\documentclass[11pt,twoside]{article}
\usepackage[dvips]{graphics}
\newcommand{\titsep}{\vspace*{0.20cm}}
\newcommand{\secsep}{\vspace*{0.50cm}}

\begin{document}
\emergencystretch=20pt
\begin{center}


{\LARGE \bf dSph Satellite Galaxies without Dark Matter: a Study of
Parameter Space\footnote{To appear in the Proceedings of the
Bonn/Bochum-Graduiertenkolleg Workshop ``The Magellanic Clouds and
Other Dwarf Galaxies'', 18.-22. January 1998, Tom Richtler \& Jochen
M. Braun (eds.), Shaker Verlag (available at
http://aibn91.astro.uni-bonn.de/~webgk/ws98/cover.html)}}

\vspace*{0.3cm}


\markboth{Pavel Kroupa}{dSph satellites without dark matter}


{\large \bf Pavel Kroupa}
\vspace*{0.3cm}


{Institut f\"ur Theoretische Astrophysik, Tiergartenstr. 15, D-69121
Heidelberg\\
e-mail: pavel@ita.uni-heidelberg.de}
\vspace*{0.85cm}

\setlength{\parindent}{0.5cm}
\hspace*{-0.6cm} \begin{minipage}{14.0cm}

A parameter study is underway in Heidelberg that aims to find and
constrain possible solutions to the dSph satellite problem under the
assumption that some of these systems may not be dark matter
dominated.  The present state of the parameter survey of tidally
disrupting spheroidal dwarf galaxies is described, and examples of
preliminary results are presented.

\end{minipage}
\end{center} 

\secsep
{\noindent \large \bf 1. Introduction} \titsep

\noindent
At least about ten dwarf spheroidal (dSph) galaxies are known to orbit
the Milky Way at distances ranging from a few tens to a few hundred
kpc. On the sky they are barely discernible stellar density
enhancements.  Their velocity dispersions are similar to those seen in
globular clusters and they have approximately the same stellar
mass. However, they are about two orders of magnitude more extended.
For spherical systems in virial equilibrium with an isotropic velocity
dispersion, the overall mass of the system can be determined from the
observed velocity dispersion.  Comparing this `gravitational' mass to
the luminosity of the system determines the mass-to-light ratio, $M/L$
(in the following always given in solar units ${\rm M}_{\odot}/{\rm
L}_{\odot}$).  Values for $M/L$ of about 10 or larger are usually
taken to imply the presence of dark matter in a stellar system.  For
the dSph satellites, $M/L$ values as large as a few hundred are
inferred, implying that these systems may be completely dark matter
dominated (for reviews see Mateo 1997, 1998).

An alternative possibility relying on Newtonian physics, is that some
of the observed dSph satellites may not be in virial equilibrium and
may not be spherical but with non-isotropic velocity dispersions, in
which case the observed $(M/L)_{\rm obs}$ ratios may not be
physical. That such systems may exist is shown by the simulations of
the long-term evolution of initially spherical satellite galaxies with
masses of $10^7\,M_\odot$, initial Plummer radii of 300~pc on
eccentric orbits in an extended Galactic dark halo with a circular
velocity of 200~km/s (Kroupa 1997, hereinafter K97).  The finding is
that after about 99~per cent of the mass is lost from the satellites
through repetitive tidal modification, remnants with $(M/L)_{\rm
obs}>10$ remain that survive for a few orbital periods. This study is
extended by Klessen \& Kroupa (1998, hereinafter KK98) covering more
orbits and Galactic halo masses using two different numerical codes.

A shortcoming common to these remnants is that they have a central
surface brightness too faint by about one order of magnitude, if
$(M/L)_{\rm true}=3$ is the mass-to-light ratio of the stellar
population.  One possible solution to this problem is to postulate
that $(M/L)_{\rm true}\approx0.3$ (K97).  However, this implies an
unusual initial mass function. Another possibility is to scale the
satellite up in mass and size so that each particle carries a larger
mass (for simulations with the same number of particles) and thus
light, while keeping the binding energy of the satellite roughly
unaffected (KK98). This, however, leads to remnants that have too
large half-light radii.

It is thus clear that the simulations mentioned above, as well as
other observational evidence summarised in K97 and KK98, give a strong
hint that some of the Galactic dSph satellites may not be dark matter
dominated but that they may be significantly affected by
tides. However, much work remains to be done to establish if models
can be produced that fit all the observational properties of at least
some of the known dSph satellites.

It is the purpose of this contribution to give a preliminary account of
the extensive parameter survey under way in Heidelberg that aims at
constraining the possible region in (initial satellite binding-energy,
mass and extend of the Galactic dark matter halo, and orbit) parameter
space in which viable ``dSph solutions'' exist in the tidal scenario.
If none such region can be found then the conclusion that all of the
dSph satellites are dark matter dominated is strengthened.

\secsep
{\noindent \large \bf 2. Simulations} \titsep

\noindent
The study of parameter space is performed with {\sc Superbox}, which
is a particle mesh code with nested sub-grids. It is described by
Madejski \& Bien (1993) with a brief description of the modern version
given by K97.  A more thorough account will be available in Fellhauer
et al. (1998). The reason for using {\sc Superbox} lies in its tremendous
efficiency: the self-consistent interaction of two galaxies consisting
in total of $1-2\times 10^6$ particles can be computed on small
workstations and personal computers with Pentium processors within a
few days.

The problem under investigation here is reduced to the interaction of
a low-mass satellite galaxy with a massive Galactic dark halo.  The
details of setting up the satellite and Galactic halo models can be
found in K97.  The dark halo is assumed to be an isothermal mass
distribution with a circular velocity of 200~km/s and a cutoff radius
of $R_{\rm c}$.  The satellite is defined by its initial mass, $M_{\rm
sat}$, and its initial Plummer radius, $R_{\rm Pl}$.  The satellite is
placed into the Galactic dark halo at apo-galactic distance $R_{\rm
apo}$ and with a tangential velocity $v_{\rm t}$. The resulting orbit
has an eccentricity $e=(R_{\rm apo}-R_{\rm peri})/(R_{\rm apo}+R_{\rm
peri})$, where $R_{\rm peri}$ is the peri-galactic distance.

The satellite is observed as an observer from Earth sees a dSph
satellite. It's projected brightness profile is measured.  The central
surface brightness and half-light radius are computed through profile
fitting, and the line-of sight velocity dispersion is determined. The
hypothetical observer thus estimates $(M/L)_{\rm obs}$ as a function
of time.

\secsep
{\noindent \large \bf 3. Some Results} \titsep

\noindent
The parameter study proceeds by varying $M_{\rm sat}$, $R_{\rm Pl}$,
$R_{\rm c}$, $R_{\rm apo}$ and $v_{\rm t}$. Most of the simulations
done so far concentrate on a satellite with $M_{\rm
sat}=10^7\,M_\odot$ and $R_{\rm Pl}=300$~pc in a Galactic dark halo
with $R_{\rm c}=250$~kpc or 40~kpc. 

Typically, $(M/L)_{\rm obs}$ remains constant at its initial low value
until {\it disruption time}, when the satellite looses most of its
remaining mass during passage through peri-galacticon.  Thereafter, a
{\it remnant} with roughly 1~per cent of the initial satellite mass
remains. It has properties rather similar to typical dSph satellites,
with large $(M/L)_{\rm obs}$.  Evolution curves for $(M/L)_{\rm
obs}(t)$ and other satellite parameters are presented in K97 and KK98.

A particularly noteworthy result obtained by KK98 is that a remnant
has a line-of-sight velocity dispersion $\sigma>6$~km/s if
$e>0.5$. This suggests a useful observational criterion, for if
$e<0.5$ for $\sigma>6$~km/s then the respective system is probably
dark matter dominated.  Here, preliminary additional results are
presented for this satellite, with some discussion of other models as
well.

\begin{figure}
\begin{center}
\rotatebox{270}{\resizebox{0.48\textwidth}{!}
{\includegraphics{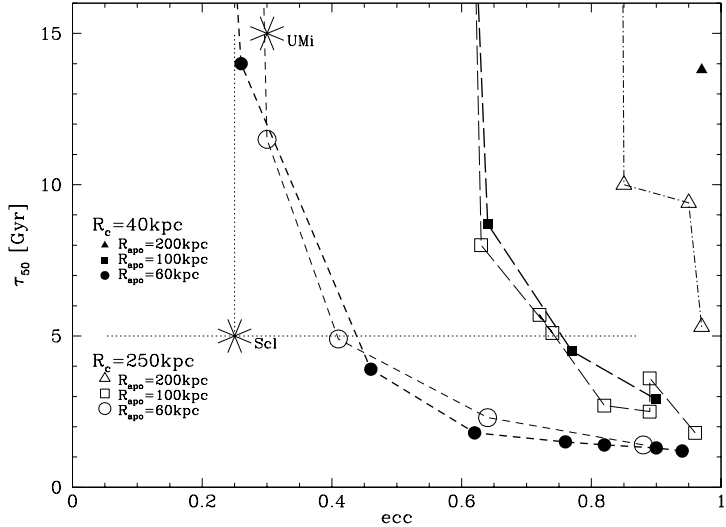}}}
\caption{The time $t=\tau_{50}$, when $(M/L)_{\rm obs}(t)\ge50$, is
plotted against orbital eccentricity for a number of simulations with
$M_{\rm sat}=10^7\,M_\odot$ and $R_{\rm Pl}=300$~pc, as defined in the
key.}
\label{fig:fig1}
\end{center}
\end{figure}

The time when $(M/L)_{\rm obs}\ge50$ occurs is denoted by $\tau_{50}$;
it increases for increasing $e$, as is evident from
Fig.~\ref{fig:fig1}.
From the figure it is apparent that $\tau_{50}$ is primarily a
function of $e$ and $R_{\rm apo}$.  Two observational constraints are
plotted: The dSph satellite {\sc Ursa Minor} contains only old stars
with ages $\tau\approx15$~Gyr (van den Bergh 1994). It's preliminary
proper motion (without an uncertainty) is reported by Olszewski
(1998), from which the orbital eccentricity is estimated.  The dSph
satellite {\sc Sculptor} has stars with ages in the range $\tau\approx
5-15$~Gyr (van den Bergh 1994), and the proper motion has been
estimated by Schweitzer et al. (1995).  Both satellites are at a
distance of 60-70~kpc. It is interesting to note that UMi has
$(M/L)_{\rm obs}=95\pm43$, while Scl has $(M/L)_{\rm obs}=10.9\pm7.5$
(Irwin \& Hatzidimitriou 1995).

The application of Fig.~\ref{fig:fig1} is as follows: If the entire
satellite was assembled a few Gyr ago then this would imply some
recent merger event strong enough to lead to the formation of tidal
tail dwarfs (e.g. Kroupa 1998). It is not clear if there is any other
evidence for such an event.  The present masses of the stellar
components (assuming $(M/L)_{\rm true}=3$) of UMi and Scl are,
respectively, about $6\times10^5\,M_\odot$ and $4\times10^6\,M_\odot$
(Irwin \& Hatzidimitriou 1995), which may be too low to accrete gas
from a hypothetical gas cloud on a similar orbit (c.f. Magellanic or
similar stream), or let alone retain gas over time-spans of
many~Gyr. Any stars with nuclear ages of a few~Gyr were probably born
from some gas-accretion event during a more massive past.  Perhaps one
may identify the nuclear age, $\tau$, of the youngest stars as being
the time when the satellite was massive enough for the last time to
accrete some gas.  The precursor satellite galaxies will have been two
orders of magnitudes more massive than the dSph satellites, if the
tidal scenario is applicable.  Thus, $\tau_{50}<\tau$ if the
respective dSph satellite is in the remnant phase with artificially
inflated $(M/L)_{\rm obs}$.

Unfortunately the observational constraints are too weak to allow any
definitive conclusions. But it is clear that if a dSph with
$(M/L)_{\rm obs}>10$ containing relatively young stars (a few Gyr old,
e.g. Scl)) can be shown with high confidence to be on a low
eccentricity ($e<0.3$) orbit, then the case for dark matter in this
system would be rather strong.

\begin{figure}
\begin{center}
\rotatebox{270}{\resizebox{0.4\textwidth}{!}
{\includegraphics{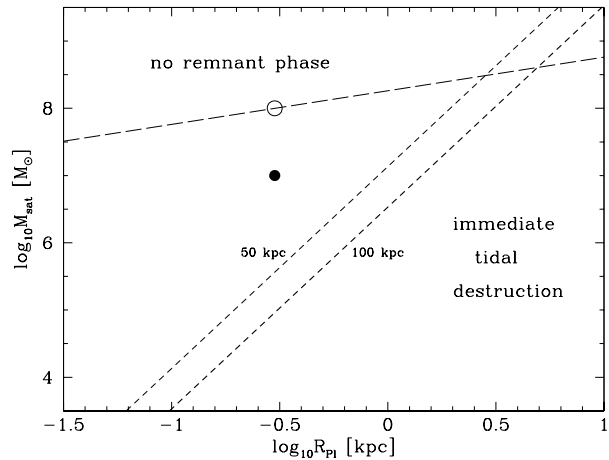}}}
\caption{Initial satellite mass as a function of initial
characteristic satellite radius. 
}
\label{fig:fig2}
\end{center}
\end{figure}

Ultimately, the parameter study will allow constraining that region in
$(M_{\rm sat},R_{\rm Pl})$ space that may lead, within a Hubble time, to
a remnant phase that agrees with the observational properties of at
least some of the dSph satellites. Fig.~\ref{fig:fig2} demonstrates
this schematically.
In this figure, the thick ($D=100$~kpc) and thin ($D=50$~kpc)
short-dashed lines are the tidal radii, log$_{10}r_{\rm
t}=-3.51+(1/3){\rm log}_{10}M_{\rm sat} + (2/3){\rm log}_{10}D$ in
kpc, where $M_{\rm sat}$ is in $M_\odot$ and $D$ is the distance from
the Galactic centre in kpc. Satellites that lie initially to the right
of these ($R_{\rm Pl}>r_{\rm t}$) are more or less immediately
destroyed on injection into the dark halo. The long-dashed line,
log$_{10}M_{\rm sat}=8.3+(1/2){\rm log}_{10}R_{\rm Pl}$ (units as
above), is the binding energy, $E_{\rm b}=(G\,M_{\rm sat}^2)/R_{\rm
pl}$, of the satellite. Satellites that initially lie above this line
do not enter the remnant phase within a Hubble time.  The filled dot
is the satellite mentioned above, and the open circle represents a
series of simulations that barely lead to any solutions, irrespective
of $e$. 

Once the solution region is delineated in this diagram, inferences can
be drawn about the physical properties of the {\it progenitors} of
those dSph satellites that are not dark matter dominated. This in turn
may allow conclusions about the physical conditions that lead to the
birth of such systems.

\secsep
{\noindent \large \bf 4. Conclusions} \titsep

\noindent
The simulations towards the parameter study briefly reported here have
been running for a couple of years, having consumed about 1~CPU~yr,
and a total of more than 13~GB of data have accumulated.  The present
results do not unambiguously demonstrate that dSph-like systems
without dark matter exist. Nevertheless, some of the models bear
surprising similarities with the dSph satellites, and some inferences
can already be drawn from this survey.

There exists in $(M_{\rm sat},R_{\rm Pl})$ space a region that leads
within a Hubble time to remnants of satellites that appear
tantalisingly similar to the dSph satellites (K97). Predictions of
correlations between observable parameters can be arrived at (KK98,
Fig.~\ref{fig:fig1}).  Also, information as to the possible initial
state of the progenitors of those dSph satellites that may not be dark
matter dominated is emerging (Fig.~\ref{fig:fig2}).

\secsep \vspace*{0.2cm}

{\noindent \large \bf References} \titsep \vspace*{-0.2cm}
\begin{description}
\itemsep=0pt \parsep=0pt \parskip=0pt \labelsep=0pt

\item Fellhauer M., Kroupa P., Bien R., et al., 1998, in preparation
\item Irwin M., Hatzidimitriou D., 1995, MNRAS, 277, 1354
\item Klessen R., Kroupa P., 1998, ApJ, in press (astro-ph/9711350)
\item Kroupa P., 1997, NewA, 2, 139, 
\item Kroupa P., 1998, in: Dynamics of Galaxies and Galactic Nuclei,
     Proc. Ser. I.T.A. no.  2, eds. W. Duschl, C. Einsel, Heidelberg
     (astro-ph/9801047)
\item Madejski R., Bien R., 1993, A\&A, 280, 383
\item Mateo M., 1997, in: The Nature of Elliptical Galaxies, 
	eds. M. Arnaboldi, G.S. Da Costa, \& P. Saha, PASP, Vol. 116 
	also: astro-ph/9701158)
\item Mateo M., 1998, these proceedings
\item Olszewski E.W., 1998, in: Proceedings of
        the 1997 Santa Cruz Halo Workshop, ed. D. Zaritsky, in
        press (astro-ph/9712280)
\item Schweitzer A.E., Cudworth
     K.M., Majewski S.R., Suntzeff N.B., 1995, AJ, 110, 2747
\item van den Bergh S., 1994, ApJ, 428, 617

\end{description}
\end{document}